# Forced Rayleigh Scattering Studies of Tracer Diffusion in a Nematic Liquid Crystal: The Relevance of Complementary Gratings

Daniel R. Spiegel, Alexis L. Thompson, and Wesley C. Campbell
Department of Physics, Trinity University, San Antonio, TX 78212-7200

**Abstract**

We have employed forced Rayleigh scattering (FRS) to study the diffusion of an azo tracer molecule (methyl red) through a nematic liquid crystal (5CB). This system was first investigated in an important study by Hara *et al*. (*Japan. J. Appl. Phys.* **23**, 1420 [1984]). Since that time, it has become clear that the presence of complementary ground-state and photoproduct FRS gratings can result in nonexponential profiles, and that complementary-grating effects are significant even when "minor" deviations from exponential decay are observed. We have investigated the methyl red/5CB system in order to evaluate the possible effects of complementary gratings. In the isotropic phase, we find that the presence of complementary gratings results in a nonmonotonic FRS signal, which significantly changes the values inferred for the isotropic diffusion coefficients. As a result, the previously reported discontinuity at the nematic/isotropic transition temperature ($T_{NI}$) is not present in the new data. On the other hand, in the nematic phase, the new experiments largely confirm the previous observations of single-exponential FRS decay and the non-Arrhenius temperature dependence of the nematic diffusion coefficients close to $T_{NI}$. Finally, we have also observed that the decrease in the diffusion anisotropy with increasing temperature can be correlated with the 5CB nematic order parameter $S(T)$ over the full nematic temperature range.

## INTRODUCTION

Many of the most fascinating properties of nematic liquid crystals (NLCs) derive from the simultaneous presence of flow and a non-zero orientational order parameter: a molecule within the NLC mesophase will undergo translational diffusion while maintaining (on average) a preferred orientation.[1] It is thus rather unfortunate that some of the most fundamental aspects of translational diffusion within NLCs are still not well understood, in spite of the impressive array of experimental, simulation, and theoretical methods that have been brought to bear on the problem.[2-16] On the experimental side, reports on NLC diffusion coefficients obtained by different methods have displayed significant discrepancies, and even results reported by different laboratories employing the same method often display marked disagreement.[4,7] The lack of consistency was severe enough to be termed an "experimental dilemma" in a recent review.[7] These difficulties have largely prevented a rigorous comparison of experimental results with fundamental theories proposed for NLC molecular diffusion,[17-20] and it is clear that experimental accuracies must be improved if such comparisons are to be realized.

One very useful experimental method for studies of diffusion in liquid crystals is forced Rayleigh scattering (FRS),[21-31] which was first employed for tracer diffusion in NLCs by Hervet *et al.* in 1978.[21] FRS is a transient grating method that is particularly well-suited for studies of NLC dynamics for three reasons. First, FRS permits real-time measurements of diffusion in the specific direction selected by the grating wavevector, which readily enables direct investigations of the diffusion anisotropy. Second, the fundamental length scale of the FRS experiment (i.e. the grating spacing, which is typically on the order of several microns) matches the intrinsic length scale at which mesomorphic fluids exhibit many of their most interesting properties.[1] Finally, FRS allows measurements at a spatial resolution determined by the diameter of the laser beams, which can easily be reduced to ~ 10$d$ (where $d$ is the grating spacing) if desired.[32,33]

In this paper we wish to consider a possible means for significant improvements in the accuracy of FRS studies on NLC systems. While mass-diffusion FRS signals were initially modeled as simple exponential functions,[21] it was later realized that an FRS signal results in general from a *difference* of two exponential decays due to "complementary" out-of-phase modulations in the ground-state and photoproduct populations.[34,35] A series of recent papers[36-38] have outlined the changes in the interpretation of FRS profiles necessitated by the presence of complementary gratings. These developments have led us to reexamine a pertinent liquid crystal system, first studied with FRS before the complementary grating model was set forth, that has displayed some surprising results. Specifically, we use FRS to investigate the azo dye molecule methyl red (MR, 2-[4-(dimethylamino)phenylazo]benzoic acid) diffusing though a cyanobiphenyl NLC (5CB, 4-$n$-pentyl-4$'$-cyanobiphenyl). The first FRS investigations on the MR/5CB system were carried out in an important study by Hara and co-workers[22] in 1984. This early work showed several intriguing features, including a non-Arrhenius temperature dependence in the nematic-phase diffusion coefficients near the nematic-isotropic transition temperature ($T_{NI}$), and an unexpected discontinuity (a positive jump) in the diffusion coefficient at $T_{NI}$. A considerably larger positive jump at $T_{NI}$ was reported for MR in MBBA in the same publication, and later papers showed positive jumps at $T_{NI}$ for MR in a number of other mesomorphic solvents, including other $n$CB compounds ($n$ = 6 - 9), 4O.8, and 8OCB.[39-42] To explain the nematic/isotropic discontinuity, it was proposed[41] that tracer diffusion through the nematic phase could be envisioned as a "dressed" process, in which the mobility of the tracer was reduced due to local distortions in the nematic continuum generated by the tracer movement, in a manner somewhat analogous to polaron motion in the solid state. The data submitted in the current report, interpreted in the context of the complementary grating model, largely confirm the important observation by Hara *et al.* of non-Arrhenius diffusion in the 5CB nematic phase; however, in the 5CB isotropic phase we observe a distinctly nonexponential decay, which significantly changes the values obtained for the isotropic diffusion coefficients. We will discuss the implications of these findings.

FRS is a tracer diffusion technique, in the sense that it measures the self-velocity correlations of detected dilute "tracer" dye molecules within a host solvent. MR has been frequently chosen as a tracer dye in liquid crystal systems due to its convenient photochromic properties and the similarity between its structure and that of liquid crystal molecules containing a two-phenyl-ring core.[21,22,25,32,39-44] However, here we make no *a priori* assumptions that tracer diffusion of MR will mimic exactly the self-diffusion of the 5CB solvent molecules, since the chemical properties of MR and 5CB are certainly different. Indeed, using FRS, Urbach *et al.*[32] and more recently Ohta and co-workers[45] have reported that MR-tracer diffusion is considerably slower than the self-diffusion of the liquid crystalline solvent molecules in nematic MBBA. It has been argued that, in NLCs, the self-diffusion of the solvent molecules and the tracer diffusion of a probe molecule of similar size will in general show the same basic features, but the anisotropy will be reduced for the latter.[4] Keeping this caveat in mind, it nonetheless remains clear that tracer diffusion is a very helpful tool for improving the understanding of NLC dynamics. FRS tracer diffusion in particular provides a very direct experimental probe of the anisotropic micron-scale dynamics of NLC fluids, which remains an area with many open questions.

**EXPERIMENTAL PROCEDURES**

MR and 5CB were purchased from Aldrich and EM Industries, respectively, and were used as received. MR/5CB solutions were prepared for FRS studies at a concentration of 0.1 wt %. The nematic-isotropic transition temperature for the 0.1% solution was found to be $(34.2 \pm 0.2)$ °C, which was not (within uncertainty) different from the measured transition temperature for the neat 5CB. The MR/5CB NLC samples were prepared by confining the mesophase between two clean microscope slides. The slides were separated by 34-gauge copper wire strands, resulting in samples with a nominal thickness of 150 µm. For studies of the temperature dependence of the diffusion coefficient in the nematic phase, the slides were first treated with an aqueous solution of poly(vinyl alcohol) (PVA) following standard procedures,[46] and then rubbed with soft paper along a specific direction to achieve planar alignment. The use of rubbed PVA coatings resulted in nearly single-domain nematic samples: at most, only a few small disclinations were observed, and these could be easily avoided when choosing the small spot (roughly 4 mm$^2$) to be examined with the FRS apparatus. For studies of the isotropic phase at temperatures greater than 36 °C, the slides were cleaned but not coated with PVA. The MR/5CB solution was sealed within the cell using a silicone adhesive.

In a FRS experiment, two interfering pump beams are used to produce a periodic fringe pattern within a sample containing a small concentration of a photochromic dye. Absorption of the pump-beam photons creates a modulation in the photoproduct and (hence) the ground-state dye population. The maxima in the ground-state population profile are coincident with the minima in the complementary photoproduct distribution. The decay of the induced modulation patterns due to diffusion can be monitored with a Bragg-diffracted probe beam. Since the ground-state and photoproduct molecules do not diffuse at identical rates, the intensity of a Bragg-diffracted homodyne-detected probe beam will decay with time according to

$$V(t) = \left(A_1 \exp[-r_1 t] - A_2 \exp[-r_2 t]\right)^2 \tag{1}$$

where the $A_i$ and the $r_i$ represent the amplitudes and decay constants, respectively, associated with the ground-state and photoproduct gratings. The negative sign arises because of the 180° spatial phase shift between the two gratings. Nearly single-exponential decay is a special case of Eq. (1), and it should be kept in mind that, according to Eq. (1), highly nonexponential (indeed, nonmonotonic) profiles are possible for an *arbitrarily small* difference between the rate constants $r_i$ if the difference between the amplitudes $A_i$ is also small.[37] Eq. (1) is strictly correct only for pure phase gratings. In the more general case of "mixed" amplitude/phase gratings, the decay of a Bragg-diffracted probe is given by[34,38]

$$V(t) = \left| A_1 \exp[-r_1 t] - e^{i\Delta f} A_2 \exp[-r_2 t] \right|^2 \qquad (2)$$

where $\Delta f$ is the difference between the optical phase shifts of the electric fields diffracted from the two gratings. The phase difference $\Delta f$ will be nonzero whenever a portion of the probe beam is absorbed within the sample.[47]

Azobenzene derivatives (such as MR) undergo a *trans* $\Rightarrow$ *cis* isomerization when excited by photons within the visible/UV absorption band. The FRS grating contrast is provided by the difference in the (complex) index of refraction between the *trans* and the *cis* species. The FRS apparatus employed for the current studies utilizes a 488-nm pump beam and a 633-nm probe beam incident at powers of 12 mW (total) and 5 mW, respectively. Since this apparatus has been described in previous reports,[36,37] the discussion here is limited to several new features relevant to studies of NLCs. The liquid-crystal sample holder consists of a machined aluminum retainer mounted to a rotation stage, which allows the director to be oriented parallel or perpendicular to the horizontal grating wavevector $q$. An 8-mm-diameter hole cut though the holder and the rotation stage allows passage of the laser beams. The stage is attached to a copper block soldered to copper pipes to permit water-flow temperature control (Neslab RTE-111). The sample temperature was calibrated in a separate run using a thermocouple junction mounted between two microscope slides and placed at the position usually occupied by the sample. The precision of the temperature control was limited to about $\pm 0.2\,^\circ\text{C}$ due to fluctuations in room temperature. The pump beam polarization was vertical (to about $\pm 1^\circ$). Following Hara *et al.*,[22] the polarization of the probe beam was rotated (using a 633-nm half-wave plate) so that the probe polarization was always parallel to the nematic director, and a spatial filter and a dichroic polarizer with its transmission axis aligned parallel to the incident probe-beam polarization were placed between the sample and the detector to reduce detection of scattered light. The probe-beam polarization was vertical in the isotropic phase. Pump beam exposure times (2 – 7 ms) were controlled using a shutter. Signals were acquired and averaged for 50 - 300 single shots on a digital storage oscilloscope (DS0). Reports in the literature[37,48] have clearly demonstrated the need for proper separation of the heterodyne and homodyne baseline components for accurate FRS results. For the present studies, the heterodyne baseline contribution (i.e., the crossterm arising from the mixing of light diffracted from the pump-induced gratings with light scattered from stationary defects) was eliminated by passing one of the pump beams through a 488-nm half-wave plate which, when rotated through $90^o$, introduces a $180^o$ shift between the phase of the probe light diffracted from the gratings and the phase of the probe light scattered from stationary defects. The sum of the two FRS profiles acquired before and after this phase shift is then a pure homodyne signal.[48] All data were acquired using this procedure.

One concern that arises with the MR/5CB system involves the potential effects of the photoproduct *cis* population on the mesoscopic properties of the NLC: since the *cis* state of MR is non-planar, photoexcitation of a MR molecule might be expected to lead to a local disruption of the nematic ordering.[45,49-52] To investigate possible unintended effects due to dye photoexcitation, we measured the FRS rate constant as a function of the pump-beam intensity in the nematic phase at a temperature of 33.9 $^o$C with $q^2 = 2.47 \times 10^8$ cm$^{-2}$. Attenuation of the pump beam by up to a factor of 14 did not result in a measurable change in the FRS rate constants for diffusion parallel or perpendicular to the director, implying that even an order-of-magnitude change in the *cis* population does not measurably affect our diffusion-coefficient results.

**RESULTS AND DISCUSSION**

FRS profiles for diffusion parallel and perpendicular to the director in the nematic phase at 33.9 °C, along with a profile for the isotropic phase at 35.6 °C, are shown in Fig. 1. The three profiles, acquired on the same day using the same PVA-coated sample, appear qualitatively similar using a DSO sensitivity in which the entire profile is acquired (Fig. 1a). If the DSO sensitivity is increased, however, the isotopic profile is seen to contain a secondary maximum, as shown in Fig. 1b. The secondary peak is quite small (less than 0.1% of the maximum signal), but is observed consistently at all grating wavevectors and temperatures investigated in the isotropic phase. Evidently, the isotropic signal is an example of a "decay-grow-decay" type of profile (with two times at which the derivative $V'(t)$ vanishes) that has appeared frequently in the FRS literature,[27,29,31,38,53,54] including at least one case of a smectic mesophase.[25]

The existence of profiles with zero, one, or two times at which the derivative $V'(t)$ is zero follows directly from Equations (1) or (2) above. Care must be exercised in fitting Equations (1) or (2) to experimental profiles via nonlinear regression, since it has been shown that such fits are in general not unique.[33,37,38] To avoid such problems, we adopted the approach recently suggested by Park *et al.*,[38] who pointed out that in many cases the *average* FRS rate constant $r_{av} = (1/2)(r_1 + r_2)$ can be determined uniquely and without difficulty from nonexponential profiles by expressing the FRS signal (Eq. [2]) in the simple form

$$V(t) \approx (at^2 + bt + c) \exp(-2 r_{av} t). \tag{3}$$

The constant coefficients *a*, *b*, and *c* can be expressed in terms of $(r_2 - r_1)$, $\Delta f$, and the $A_i$. Equation (3) is obtained from a Taylor series expansion of Eq. (2) about the time $t_0 = (r_2 - r_1)^{-1} \{\ln(A_2/A_1) + i\Delta f\}$, and represents a useful approximation for nonmonotonic profiles whenever the fractional differences $(A_2 - A_1)/A_1$ and $(r_2 - r_1)/r_1$ are not very large (as expected for azo dyes),[35,55] and $\Delta f \ll 1$ rad. Equation (3) has four unknown parameters (one less than Eq. [2]), three of which appear linearly; thus, Eq. (3) is expected to be far less susceptible to the problems with non-unique fits reported using Equations (1) or (2).[27,33,38,56] For pure phase gratings ($\Delta f = 0$), in which case $t_0$ is real and represents the time at which the signal $V(t)$ is zero, Eq. (3) reduces to

$$V(t) \approx a'(t - t_0)^2 \exp(-2 r_{av} t) \tag{4}$$

which contains three unknown parameters (one less than Eq. [1]). In Fig. 2 it can be seen that the minimum in the isotropic profile does not reach the baseline, which demonstrates that $\Delta f \neq 0$ due to a small amount of probe beam absorption within the long-wavelength tail of the visible MR absorption band. We therefore fit all isotropic profiles to Eq. (3) to obtain $r_{av}$. An example of a fit to Eq. (3) is also shown in Fig. 2.

Turning to the nematic profiles, it is important to determine if single-exponential fits are appropriate in this case. It has been shown that,[37] when the photoproduct and ground-state rate constants differ by a small amount, forcing a single-exponential fit to a monotonic FRS profile can lead to significant errors, since the semi-log slope $r_{exp} = (-1/2)\, d(\ln[V])/dt$ does not represent a physically useful FRS rate constant unless the dimensionless curvature $K_2/(K_1)^2$ is zero within experimental error. $K_1$ and $K_2$ are the first and second derivatives of $(1/2)\ln(V[t])$ at $t = 0$ and can be used to provide estimates of the first and second cumulants, respectively, of the FRS decay.[57] We applied second-order polynomial fits to multiple profiles to obtain averages of $K_2/(K_1)^2$ at two different temperatures with the director orientated either perpendicular or parallel to the grating wavevector. In each case, four signal profiles obtained at different grating wavevectors were each fit over the time interval for which $V(t)$ varied from 90% down to 40% of its maximum value. (The top 10% of the profile was not included in the fit to allow time for the much faster thermal-grating decay.)[37] At a temperature of 33.8 °C, we obtained, using ±1σ uncertainties, $\langle K_2/(K_1)^2 \rangle_\perp = 0.13 \pm 0.09$ and $\langle K_2/(K_1)^2 \rangle_\| = -0.1 \pm 0.1$. At a temperature of 25.9 °C, we measured $\langle K_2/(K_1)^2 \rangle_\perp = -0.08 \pm 0.08$ and $\langle K_2/(K_1)^2 \rangle_\| = 0.003 \pm 0.07$. Thus the curvature ratio is zero within uncertainty, which validates quantitatively the application of single-

exponential fits to the nematic data and implies that the semi-log slope is equal (within uncertainty) to the rate constant $r_{av}$.[37] All nematic rate constants were then obtained using single-exponential fits over the time interval for which $V(t)$ varied from 90% down to 2% of its maximum value.

Examples of rate constants obtained from single-exponential fits to the nematic profiles, and from fits to Eq. (3) for the isotropic profiles, are shown as a function of $q^2$ in Fig. 3. All diffusion coefficients in this report were obtained from the slope of such plots using 4 different values of $q^2$. In Fig. 4 we show the diffusion coefficient as a function of temperature for diffusion parallel and perpendicular to the director in the nematic phase, along with the diffusion coefficient in the isotropic phase. From sample-to-sample variations in the rate constants we estimate the maximum error in the diffusion coefficient to be ±3% in the nematic phase and ±6% in the isotropic phase. The temperature dependence is arguably of the Arrhenius type for the isotropic phase and in the low-temperature ($T < 30\ ^\circ C$) region of the nematic phase, although certainly any assignment of an "Arrhenius" dependence over the rather narrow temperature regions studied must be made with due caution. Table I reports the corresponding isotropic activation energy $\Delta E_{iso}$ and the nematic-phase activation energies $\Delta E_\parallel$ and $\Delta E_\perp$. There is no obvious break in the temperature dependence at the nematic-isotropic transition temperature: $D_\parallel$ and $D_{iso}$ are essentially continuous across $T_{NI}$.

As the temperature approaches $T_{NI}$ in the nematic phase, the behavior is clearly non-Arrhenius. Although both $D_\parallel$ and $D_\perp$ display a decreasing slope as $T \rightarrow T_{NI}$, the effect is more dramatic for $D_{||}$, so that the anisotropy ratio $D_{||}/D_{\perp}$ decreases as the phase transition is approached. The variation in $D_{||}/D_{\perp}$ can be exhibited in a more direct fashion by plotting this ratio as a function of the nematic order parameter $S(T)$. To obtain the latter, we use $S(T)$ as derived by Sherrell and Crellin[58] from their measurements of the anisotropy in the 5CB nematic-phase magnetic susceptibility, which were later recommended by Ahlers[59] as the best available values. $D_{||}/D_{\perp}$ is plotted as a function of $S(T)$ for the full nematic temperature range in the inset of Fig. 4. As expected from first principles,[18,19] the diffusion anisotropy measured using FRS increases as the 5CB nematic order parameter increases, although improved precision will certainly be necessary before the correct quantitative relation between these two quantities can be identified with confidence.

We may now compare the current work to a previous study by Hara et al.,[22] who also carried out FRS studies on 0.1% MR-5CB samples confined between rubbed PVA-treated slides. As in the current study, rate constants in the nematic phase were obtained using single-exponential fits. For comparison, the nematic activation energies and the values of $D_{||}$ and $D_{\perp}$ at 25 °C obtained by Hara et al. are included along with the results of the present work in Table I. There is good agreement in the values of the diffusion coefficients at 25 °C. On the other hand, Hara et al. obtained activation energies for parallel and perpendicular diffusion coefficients that differed by about 80%, while the two activation energies are nearly equal for the current investigations. It is worth noting, however, that the scatter in the diffusion coefficient vs. temperature measurement of Hara et al. is larger than that of the present study, and that the average of the parallel and perpendicular activation energies reported previously is close to the average obtained in the current study. Hara *et al.* carefully noted a decrease in the slope of $D_{||}$ as the temperature approached $T_{NI}$, and Figure 6 of their paper shows a possible decrease in the slope of $D_\perp$ in this temperature region. The present results largely confirm these important observations (at a somewhat higher internal precision), and also show that the non-Arrhenius dependence can be correlated with the nematic order parameter.

In their studies of the isotropic MR/5CB phase, Hara *et al.* also fit the FRS profiles to single-exponential decays. The secondary peak shown in Fig. 1b of the current report was apparently not detected, presumably because of its small amplitude relative to the maximum signal level. (It should be noted that there was no reason to expect a secondary maximum when the Hara *et al.* work was carried out, since the significance of complementary FRS gratings had not yet been identified.[34,35]) The data of Hara *et al.* show (in their words) a "discontinuous jump" in the diffusion coefficient at $T_{NI}$, with $D_{iso}(T_{NI})$ exceeding $D_{||}(T_{NI})$ by ~ 30%. The current studies treat the FRS signals in the context of the complementary grating model, yielding a smaller value for $D_{iso}$ at $T_{NI}$ that is (within error) nearly continuous with $D_{||}$, along with an isotropic activation energy $\Delta E_{iso}$ equal to about 2/3 of the value obtained previously (see Table I). Thus it is apparent that the unexpected discontinuity at $T_{NI}$ is not observed when the complementary grating characteristics of the FRS signals are accounted for. It is not surprising that the isotropic diffusion coefficients obtained in the previous investigation are consistently larger than those measured in the current study, since it is easily shown that, for nearly-pure phase gratings, the rate constant $r_{exp} = (-1/2)\, d(\ln[V])/dt$ extracted by fitting a decay-grow-decay FRS profile to a single exponential function will always be larger than either $r_1$ or $r_2$.[37] The rather unremarkable transition from nematic to isotropic diffusion evident in Fig. 4 implies that it is not necessary to invoke a hypothesis involving "dressed" polaron-like tracer diffusion within the 5CB nematic phase. As noted above, nematic/isotropic diffusion discontinuities have been reported in FRS studies for MR in other NLC systems.[39-42] Although the current experiment involves only 5CB, it is possible that the nematic/isotropic discontinuities reported for these other systems could also be related to complementary grating effects. In any case, it is clear that complementary grating considerations are important for accurate FRS studies of dynamics within liquid crystals.

**CONCLUSIONS**

New FRS studies carried out using the complementary grating model to interpret data obtained at higher sensitivity confirm the non-Arrhenius behavior reported previously for MR in nematic 5CB. We have also found that the diffusion anisotropy ratio $D_\parallel/D_\perp$ can be correlated with the nematic order parameter $S(T)$ over the full nematic temperature range. In the isotropic phase, we have observed a decay-grow-decay FRS profile, resulting in new values for the isotopic diffusion coefficient and its activation energy that are significantly lower than the previous values; this in turn removes the previously reported diffusion anomaly at the nematic-isotropic phase transition. Our results highlight the importance of accounting for complementary grating effects for accurate FRS measurements in liquid-crystal systems.

**ACKNOWLEDGMENTS**

We gratefully acknowledge Dr. Taihyun Chang for his careful critique of the manuscript, and we thank Dr. Thomas Moses and Dr. Benjamin Plummer for very useful discussions. We are grateful to Mr. Tom Defayette for his extraordinary dedication and expertise in the machine shop and the laboratory, to Mr. Richard Helmer for his highly valuable contributions to the electronics, and to Mr. Timothy Gilheart and Mr. Dustin Ragan for their assistance. Finally, we express our appreciation to the anonymous reviewer, whose careful critique resulted in (we believe) significant improvements to the clarity of the manuscript. The Trinity forced Rayleigh scattering laboratory is funded by a grant from the National Science Foundation (RUI CHE-9711426).

**References**

1 P. G. de Gennes and J. Prost, *The Physics of Liquid Crystals (2nd edition)* (Clarendon Press, Oxford, 1993).

2 C. K. Yun and A. G. Fredrickson, *Mol. Cryst. and Liq. Cryst.* **12,** 73 (1970).

3 A. J. Leadbetter, F. P. Temme, A. Heidemann, and W. S. Howells, *Chem. Phys. Lett.* **34,** 363 (1975).

4 G. J. Krüger, *Phys. Rep.* **82,** 229 (1982).

5 A. G. Chmielewski, *Mol. Cryst. Liq. Cryst.* **212,** 205 (1992).

6 C. W. Cross and B. M. Fung, *J. Chem. Phys.* **101,** 6839 (1994).

7 F. Noack, in *Handbook of Liquid Crystals (Vol. 1: Fundamentals)*, edited by D. Demus, J. Goodby, G. W. Gray, H.-W. Spiess, and V. Vill (Wiley-VCH, Weinheim, 1998).

8 P. Holstein, M. Bender, P. Galvosas, D. Geschke, and J. Kärger, *J. Magn. Reson.* **143,** 427 (2000).

9 S. V. Dvinskikh, R. Sitnikov, and I. Furó, *J. Magn. Reson.* **142,** 102 (2000).

10 P. Etchegoin, *Phys. Rev. E* **59,** 1860 (1999).

11 C. M. Snively and J. L. Koenig, *J. Polym. Sci. B: Polym. Phys.* **37,** 2261 (1999).

12 D. Wiersma, A. Muzzi, M. Colocci, and R. Righini, *Phys. Rev. Lett.* **83,** 4321 (1999).

13 N. Nakajima, N. Hirota, and M. Terazima, *J. Photochem. Photobiol. A: Chemistry* **120,** 1 (1999).

14 M. P. B. van Bruggen, H. N. W. Lekkerkerker, G. Maret, and J. K. G. Dhont, *Phys. Rev. E* **58,** 7668 (1998).

15 A. A. Khare, D. A. Kofke, and G. T. Evans, *Mol. Phys.* **91,** 993 (1997).

16 S. Ravichandran, A. Perera, M. Moreau, and B. Bagchi, *J. Chem. Phys.* **107,** 8469 (1997).

17 W. Franklin, *Phys. Rev. A* **11,** 2156 (1975).

18 K. -S. Chu and D. S. Moroi, *J. de Physique* **36,** C1-99 (1975).

19 S. Hess, D. Frenkel, and M. P. Allen, *Mol. Phys.* **74,** 765 (1991).

20 D. Sokolowska and J. K. Moscicki, *Phys. Rev. E* **54,** 5221 (1996).

21 H. Hervet, W. Urbach, and F. Rondelez, *J. Chem. Phys.* **68,** 2725 (1978).

22 M. Hara, S. Ichikawa, H. Takezoe, and A. Fukuda, *Japan. J. Appl. Phys.* **23,** 1420 (1984).

23 P. Fabre, L. Léger, and M. Veyssie, *Phys. Rev. Lett.* **59,** 210 (1987).

24 L. Wang, M. M. Garner, and H. Yu, *Macromolecules* **24,** 2368 (1991).

25 T. Moriyama, Y. Takanishi, K. Ishikawa, H. Takezoe, and A. Fukuda, *Liq. Cryst.* **18,** 639 (1995).

26 M. Terazima, Y. Kojima, and N. Hirota, *Chem. Phys. Lett.* **259,** 451 (1996).

27 T. Lodge and B. Chapman, *Trends Polym. Sci.* **5,** 122 (1997).

28 B. Yoon, S. H. Kim, I. Lee, S. K. Kim, M. Cho, and H. Kim, *J. Phys. Chem. B* **102,** 7705 (1998).

29 J. Xia and C. H. Wang, *Macromolecules* **32,** 5655 (1999).

30 A. V. Veniaminov and H. Sillescu, *Chem. Phys. Lett.* **303,** 499 (1999).

31 C. Graf, W. Schärtl, M. Maskos, and M. Schmidt, *J. Chem. Phys.* **112,** 3031 (2000).

32 W. Urbach, H. Hervet, and F. Rondelez, *J. Chem. Phys.* **83,** 1877 (1985).

[33] W. J. Huang, T. S. Frick, M. R. Landry, J. A. Lee, T. P. Lodge, and M. Tirrell, *AIChE J.* **33,** 573 (1987).

[34] C. S. Johnson, *J. Opt. Soc. Am. B* **2,** 317 (1985).

[35] S. Park, J. Sung, H. Kim, and T. Chang, *J. Phys. Chem.* **95,** 7121 (1991).

[36] D. R. Spiegel, M. B. Sprinkle, and T. Chang, *J. Chem. Phys.* **104,** 4920 (1996).

[37] D. R. Spiegel, A. H. Marshall, N. T. Jukam, H. S. Park, and T. Chang, *J. Chem. Phys.* **109,** 267 (1998).

[38] H. S. Park, T. Chang, and D. R. Spiegel, *J. Chem. Phys.* **112,** 9518 (2000).

[39] M. Hara, H. Tenmei, S. Ichikawa, H. Takezoe, and A. Fukuda, *Japan. J. Appl. Phys.* **24,** L777 (1985).

[40] H. Takezoe, M. Hara, S. Ichikawa, and A. Fukuda, *Mol. Cryst. Liq. Cryst.* **122,** 169 (1985).

[41] M. Hara, H. Takezoe, and A. Fukuda, *Japan. J. Appl. Phys.* **25,** 1756 (1986).

[42] T. Nishikawa, J. Minabe, H. Takezoe, and A. Fukuda, *Mol. Cryst. Liq. Cryst.* **231,** 153 (1993).

[43] H. Takezoe, S. Ichikawa, A. Fukuda, and E. Kuze, *Japan. J. App. Phys.* **23,** L78 (1984).

[44] D. Jin, H. Kim, S. H. Kim, and S. K. Kim, *J. Phys. Chem. B* **101,** 10757 (1997).

[45] K. Ohta, M. Terazima, and N. Hirota, *Bull. Chem. Soc. Jpn.* **68,** 2809 (1995).

[46] T. Uchida and H. Seki, in *Liquid Crystals: Applications and Uses (Vol. 3)*, edited by B. Bahadur (World Scientific, Singapore, 1990).

[47] H. Kogelnik, *Bell Syst. Tech. J.* **48,** 2909 (1969).

[48] W. Köhler, *J. Chem. Phys.* **98,** 660 (1993).

[49] I. Jánossy, *Phys. Rev. E* **49,** 2957 (1994).

[50] R. Muenster, M. Jarasch, X. Zhuang, and Y. R. Shen, *Phys. Rev. Lett.* **78,** 42 (1997).

[51] T. V. Galstian, B. Saâd, and M. -M. Denariez-Roberge, *IEEE J. Quantum Electron.* **34,** 790 (1998).

[52] M. Nöllmann, D. Shalóm, P. Etchegoin, and J. Sereni, *Phys. Rev. E* **59,** 1850 (1999).

[53] M. Terazima, *Chem. Phys. Lett.* **304,** 343 (1999).

[54] K. Gohr and W. Schärtl, *Macromolecules* **33,** 2129 (2000).

[55] L. S. Lever, M. S. Bradley, and C. S. Johnson, *J. Magn. Reson.* **68,** 335 (1986).

[56] B. R. Chapman, C. R. Gochanour, and M. E. Paulaitis, *Macromolecules* **29,** 5635 (1996).

[57] D. E. Koppel, *J. Chem. Phys.* **57,** 4814 (1972).

[58] P. L. Sherrell and D. A. Crellin, *J. de Physique* **40,** C3-211 (1979).

[59] G. Ahlers, in *Pattern Formation in Liquid Crystals*, edited by A. Buka and L. Kramer (Springer, New York, 1995).

**Table I. Diffusion of Methyl Red in 5CB.** For the present work, activation energies in the nematic phase are obtained for $T < 30\,^\circ\text{C}$ (see Fig. 4). The results of Hara *et al.* were published in *Japan. J. Appl. Phys.* **23,** 1420 (1984).

|  | Diffusion Constant ($\times 10^{-7}$ cm$^2$/s) | | Activation Energy (kJ/mol) | | |
|---|---|---|---|---|---|
|  | $D_\parallel(25\,^\circ\text{C})$ | $D_\perp(25\,^\circ\text{C})$ | $\Delta E_\parallel$ | $\Delta E_\perp$ | $\Delta E_{\text{iso}}$ |
| Present work | $2.3 \pm 0.1$ | $1.5 \pm 0.1$ | $28 \pm 1$ | $31 \pm 2$ | $36.1 \pm 0.3$ |
| Hara *et al.* | 2.3 | 1.4 | 22 | 40 | 54 |

**Figure Captions**

**Fig. 1.** FRS signals obtained in MR/5CB samples at $q^2 = 5.50 \times 10^7$ cm$^{-2}$. Profiles are displayed with the director oriented perpendicular ($\perp$) and parallel ($\parallel$) to the grating wavevector in the nematic phase at 33.9 °C, along with a profile in the isotropic ("iso") phase at 35.6 °C. (*a*) Full profiles, with horizontal and vertical offsets added for clarity. To avoid detector saturation, the perpendicular and parallel nematic signals were attenuated by 75% and 30%, respectively, using neutral density filters placed between the sample and the detector. (*b*) The signal obtained at an increased oscilloscope sensitivity (1 mV/division), with no additional changes. Horizontal and vertical offsets have been added for clarity. It is apparent that the isotropic profile is nonmonotonic. (*c*) Logarithm of the signals shown in parts (*a*) and (*b*). Vertical offsets have been added to the logarithmic profiles for clarity. The smaller-size data points were obtained at the higher oscilloscope sensitivity.

**Fig. 2.** A FRS profile obtained in the isotropic phase at a temperature of 34.8 °C using a grating wavevector with $q^2 = 7.29 \times 10^7$ cm$^{-2}$. For clarity, only 25% (every fourth point) of the data have been plotted. A fit to Eq. (3) (solid line) yields $a = 229$ V/s$^2$, $b = -48.9$ V/s, $c = 2.61$ V, and $r_{av} = 23.7$ Hz. The residuals (with every fourth point plotted) are shown in the inset.

**Fig. 3.** Plots of the rate constant as a function of the square of the grating wavevector. In the upper plot, the nematic rate constants, measured with the director oriented parallel ($\parallel$) and perpendicular ($\perp$) to the grating wavevector, were obtained from simple exponential fits. In the lower plot, the rate constants in the isotropic phase were obtained from fits to Eq. (3).

**Fig. 4.** Tracer diffusion coefficients for MR in 5CB as a function of temperature in the nematic and isotropic phases. The inset shows the ratio $D_{\parallel}/D_{\perp}$ as a function of the 5CB nematic order parameter $S(T)$ for the full nematic temperature range.

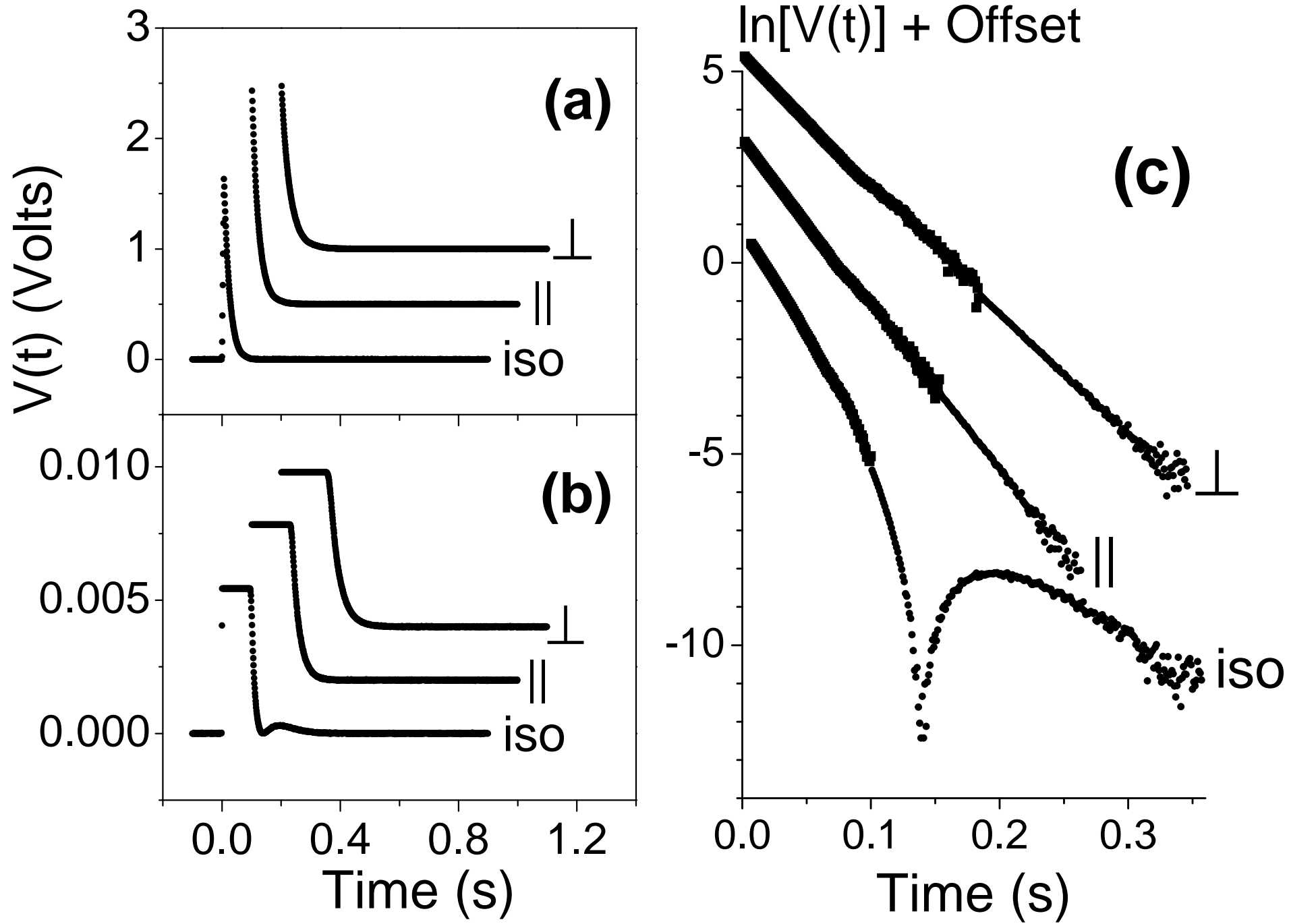

Fig. 1 Spiegel et al. JCP

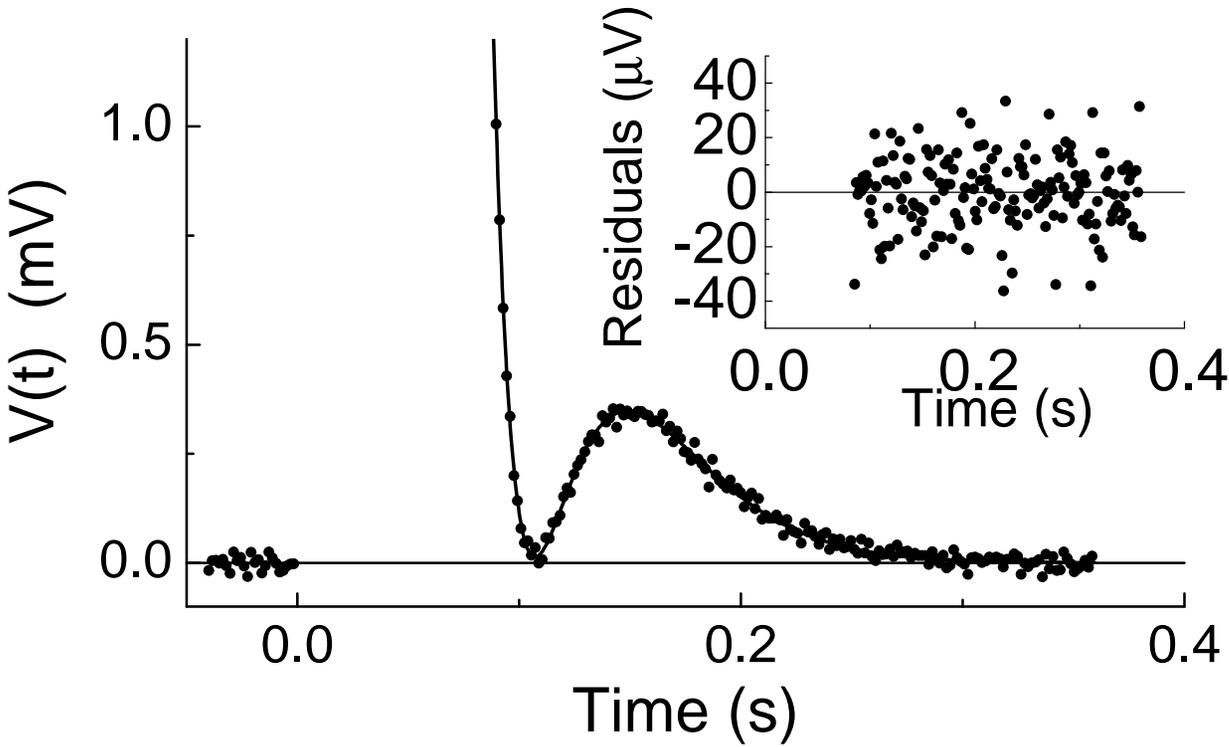

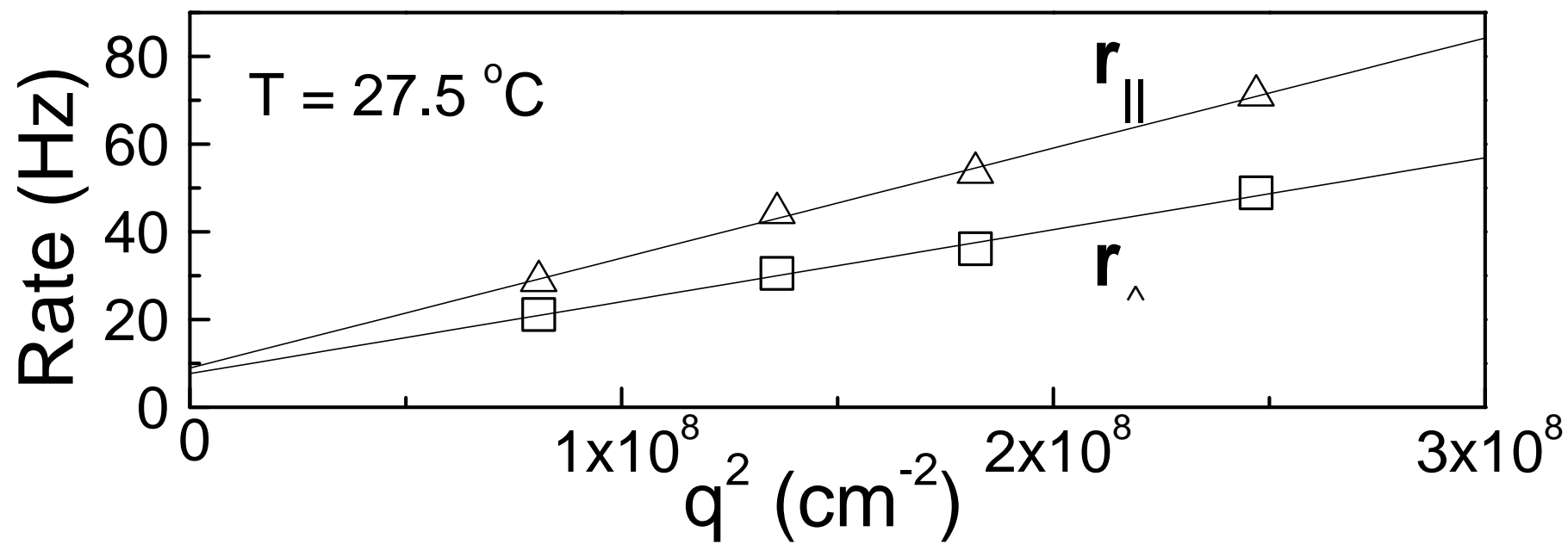
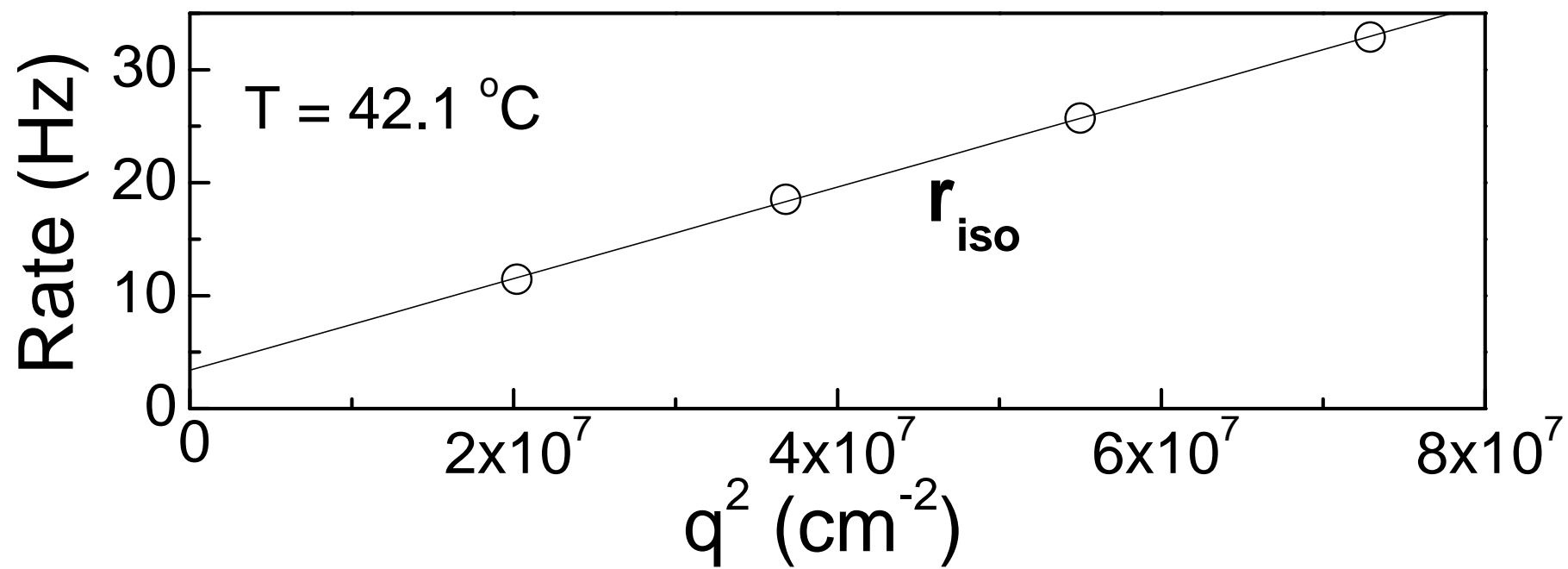

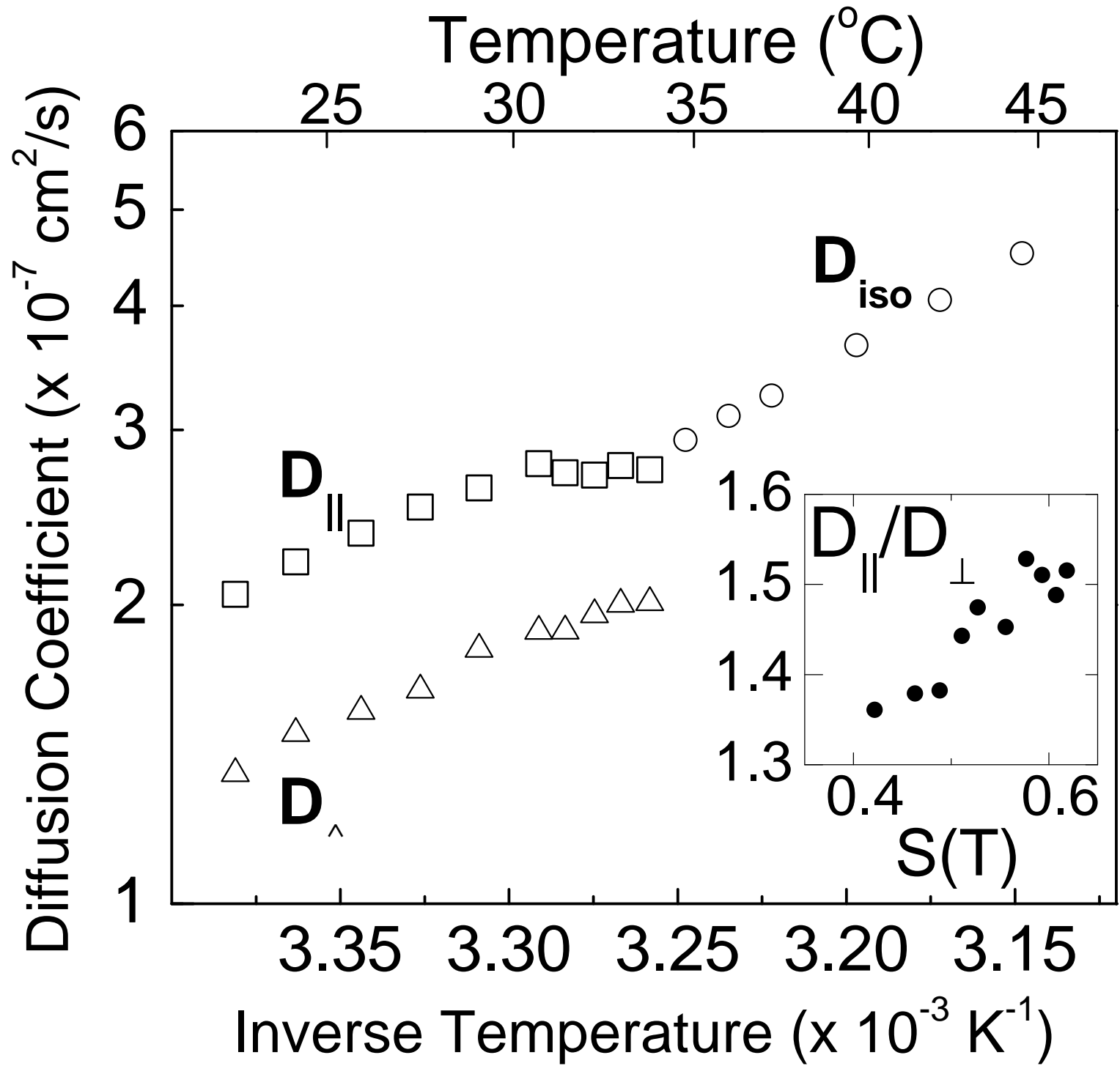

Fig. 4 Spiegel et al. JCP